\documentclass[prl,final,twocolumn,showpacs,byrevtex,float]{revtex4}
\usepackage{graphicx}
\usepackage{epsfig}
\begin{document}
\def\BibTeX{\rm B{\sc ib}\TeX}


\title{High Van Hove singularity extension and Fermi velocity increase in epitaxial graphene functionalized by gold clusters intercalation\\}
\author{M. N. Nair$^{1}$, M. Cranney$^{1}$, F. Vonau$^{1}$, D. Aubel$^{1}$, P. Le F\`{e}vre$^{2}$, A. Tejeda$^{2,3}$, F. Bertran$^{2}$, A. Taleb-Ibrahimi$^{2}$
and L. Simon$^{1}$$\footnote[1]{corresponding author \\Email
address: laurent.simon@uha.fr}$} $^{1}$
\affiliation{$^{1}$Institut de Sciences des Mat\'{e}riaux de
Mulhouse IS2M-LRC 7228-CNRS-UHA 4, rue des fr\`eres Lumi\`ere
68093 Mulhouse-France}
\affiliation{$^{2}$Synchrotron SOLEIL, L'Orme des Merisiers, Saint-Aubin, 91192 Gif sur Yvette, France}\\
\affiliation{$^{3}$Institut Jean Lamour, CNRS-Universit\'{e} de
Nancy-UPV-Metz, 54506 Vandoeuvre les Nancy, France}\\
\date{\today}

\begin{abstract}
Gold intercalation between the buffer layer and a graphene
monolayer of epitaxial graphene on SiC(0001) leads to the
formation of quasi free standing small aggregates of clusters.
Angle Resolved Photoemission Spectroscopy measurements reveal that
these clusters preserve the linear dispersion of the graphene
quasiparticles and surprisingly increase their Fermi velocity.
They also strongly modify the band structure of graphene around
the Van Hove singularities (VHs) by a strong extension without
charge transfer. This result gives a new insight on the role of
the intercalant in the renormalization of the bare electronic band
structure of graphene usually observed in Graphite and Graphene
Intercalation Compounds.

\end{abstract}

\pacs{68.65.-k, 81.16.Fg, 81.07.-b, 81.16.Rf, 82.30.RS, 82.65.+r}

\maketitle

The enormous craze for graphene is due to the coexistence between
the fundamental aspects of research and the increasing number of
potential applications. From the  fundamental point of view, this
system brings together the physics of particles with relativistic
behaviors and the condensed matter. It becomes a promising
material for the next generation of nanoelectronic devices
destined to supplant silicon \cite{NovoselovNature2005}. However
the main drawback limiting the potential use of graphene stems
from its intrinsic characteristics: a semiconductor with zero gap,
almost inert towards controlled chemisorption and doping. One of
the main challenges is to functionalize the graphene layer while
preserving its fascinating properties. Different ways of
functionalization have been opened. Deposition of metal or
molecules on top of graphene could allow to modify the Fermi level
or to induce long range superconductive correlations (for example
using superconducting metal contact)
\cite{BouchiatPRL2010,DirksNatPhysics2011}. It is also possible to
intercalate metal clusters or molecules between the graphene
layers, opening the possibility to functionalize the graphene
layer on both sides creating a Graphene-Based Hybrid structure
(GBHs). In the latter case,  the potentialities of modification of
the graphene band structure by intercalation meet the historical
research community of the Graphite Intercalation Compounds (GICs),
well-known in the community of Carbon and also for its famous
application which is the Li-ion battery. The research in this
field has been considerably intensified after the recent discovery
of high Tc superconductivity for the GIC $CaC_{6}$
\cite{WellerNatPhys2005}. Despite  this intense activity, it is
still not yet clear if the superconductivity is due to the nature
of the intercalant or to the graphene plane itself. Angle Resolved
Photoemission Spectroscopy (ARPES)  measurements have revealed a
systematic VHs extension for these graphitic superconductor
\cite{VallaPRL2011}. Moreover, recently, in the case of $CaC_{6}$
the superconducting property has been associated to Charge Density
Waves (CDW) evidenced by STM \cite{EllerbyNature2011}. In this
context, epitaxial graphene consists of a playground to understand
how the band structure of graphene could be modified and more
particularly the various many-body phases that we could expect
near VHs. Indeed, the growth of graphene monolayer on silicon face
of silicon carbide substrate leads to the formation of a monolayer
graphene covalently bonded to the substrate (called buffer layer)
which decouples the true monolayer graphene in weak interaction
with it. An intercalation process is then possible between these
two layers. As the transition from the monolayer to bilayer and
few layers graphene can be done in a controlled way on SiC(0001),
this system is also particularly interesting to follow  the
staging sequences of the intercalation process. In a detailed
study by Scanning Tunneling Microscopy (STM), we have revealed
that upon specific preparation procedure gold intercalates in two
different structures \cite{PremlalAPL09}. One is the formation of
small intercalated clusters. We have shown by STM that these
intercalated gold clusters create a  strong "standing waves-like"
pattern, on the upper monolayer graphene which has been attributed
to a possible VHs singularity extension \cite{cranneyEPL10}.
\\The goal of the present work was to realize homogeneous surfaces with intercalated gold clusters
and to explore the band structure of this functionalized graphene
on the occupied states  with ARPES measurements. We report here
ARPES and STM studies of pristine epitaxial graphene  and
with the intercalation of gold clusters.\\

The graphene samples were prepared in UHV by the annealing of
n-doped SiC(0001) at 900 K for several hours and subsequent
annealing at 1500 K \cite{VanBommelSurfSci75, SimonPRB99,others}.
The deposition of gold on graphene was carried out at room
temperature using a homemade Knudsen cell calibrated using a
Quartz Crystal Microbalance. The sample was further annealed at
1000 K for 5 min \cite{PremlalAPL09}. In order to avoid frequent
confusion, we would like to notice that the literature reports two
types of intercalation process. One occurs during the annealing
process of the carbon rich SiC(0001) reconstruction in presence of
a foreign element for example H, F or Au
\cite{BostwickScience2010,
WalterAPL2011,GierzPRB2010,StarkePRB11,WongACSNano11}. This leads
to the intercalation between the SiC substrate and the C-rich
initially covalently bonded graphene layer leading to a partial
decoupling. This is associated to a p-type doping effect. In our
case the deposition of gold is done after the complete realization
of the graphene monolayer leading to the intercalation between the
top graphene layer and the buffer layer. Our STM experiments were
performed with a LT-STM from Omicron at 77 K at a base pressure in
the $10^{-11}$ mbar range. The $dI/dV$ images were acquired using
a lock-in amplifier and a modulation voltage of $\pm 20 mV$. The
ARPES measurements were carried out on the CASSIOPEE beamline of
the SOLEIL synchrotron radiation source using a Scienta R4000
electron spectrometer. The spectra were recorded at a photon
energy of 60 eV with an overall energy resolution of around 30 meV
at a temperature of 10 K. All the samples were prepared in a
separated UHV system and then characterized  by STM. They were
then transported in air prior to their introduction in the UHV
system of the synchrotron radiation source. A soft degassing
process at 500K was performed during several minutes prior to
measurements.
\begin{figure}
\includegraphics[width=8cm]{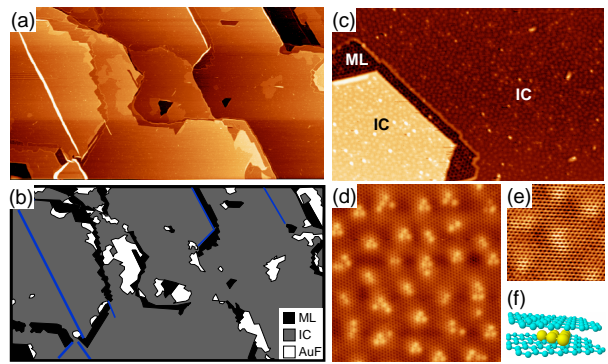}
\caption{STM pictures at 77K of the surface of epitaxial graphene
obtained after 18 ML gold deposition followed by 5 min of
annealing at 1000 K. In (a) a large-area topographic image
(500$\times$276 $nm^{2}$, -1.5 V) and the corresponding scheme (b)
below show the repartition of the different domains, where the
pristine monolayer graphene domain (ML) is in black, intercalated
clusters domain (IC) in grey and gold film domain (AuF) in white
(the puckers are in blue). (c) shows the pristine monolayer
graphene (ML) and the intercalated clusters region (IC) on two
different terraces. The IC region consists of the intercalation of
aggregates of flat clusters between the first monolayer graphene
and the buffer layer as zoomed in (d) and as schematized in (f).
(e) The STM image ($5.6 \times 5.6$ $nm^{2}$, -100 mV) ascertains
that Au clusters are intercalated below a monolayer graphene; (c)
($111\times70$ $nm^{2}$, -1.5 V) and (d) ($111\times70$ $nm^{2}$,
-1.5 V). (Image processing using the WSxM software \cite{wsxm}).}
\label{Fig1}\end{figure}

Figure \ref{Fig1} shows the resulting STM images of the epitaxial
graphene monolayer with the deposition of gold atoms followed by
annealing process as previously described. ARPES measurements
require surfaces as homogeneous as possible, which has been done
here. We are able to control the deposition and annealing process
in order to obtain a  fairly homogeneous surface  with the given
intercalation process. Indeed figures \ref{Fig1}a) and b) show
respectively a large scale STM image of the studied sample and the
corresponding scheme of the repartition of the different domains
obtained. More than 80\% of the surface is covered with the
intercalated gold clusters (IC) . A small part is occupied by
pristine monolayer (ML) and by another domain which corresponds to
the insertion of a continuous monolayer of gold (AuF). The
continuous monolayer of gold has been evidenced by  a Moir\'{e}
pattern and is associated to a p-doping effect
\cite{PremlalAPL09}. The bright lines in a) correspond to the
initial puckers currently observed on the pristine  ML graphene,
that are known to be due to the cooling process after the
annealing. These defects play probably a role in the intercalation
of metal as they are systematically observed on the border of
areas with intercalated gold. As shown in figures c) and d), the
gold clusters are evidenced by bright protrusion visible at high
negative bias (probing the full states). The intercalated gold
atoms on the IC domains form a quasi periodically arrangement of
aggregates of clusters intercalated between the buffer layer and
the top graphene monolayer. Indeed in e) a high resolution image
performed at low bias (-100 meV) shows the graphene plane over
gold clusters where the 6 carbon atoms of the honeycomb structure
are equally visible. This definitively proves that gold clusters
are just under the top graphene monolayer. These clusters are less
visible for the low bias voltage and the honeycomb structure of
graphene dominates the contrast. In Figure d), we tentatively
attributed these bright spots to aggregates of flat clusters made
of 6 atoms as schematized in figure f) \cite{PremlalAPL09}. This
proposed model is however still under debate and currently tested
by DFT calculations, but the discussion of the exact nature of
these
clusters is out of the scope of this article.\\

We have shown that these clusters create standing waves  patterns
for bias voltage corresponding to the unoccupied states starting
from +0.6 to +1 eV
 \cite{cranneyEPL10}. The non-dispersive character of these
standing waves let us consider them as charge density waves (CDW).
We have attributed these structures as a screening effect. Indeed
we know that in the epitaxial graphene the top Graphene ML is
n-doped due to the transfer of charges from the substrate. The Au
clusters seem to screen these charges which create a charge
inhomogeneity on the graphene plane and scatter the QPs. We have
used the Fourier Transform Scanning Tunnelling Spectroscopy
Technique (FT-STS) \cite{JofCondMater07}  i. e. we performed a 2D
FT of the dI/dV map images with standing waves pattern in function
of the bias voltage. This technique was already successfully used
in order to determine the full band structure dispersion of the 2D
$ErSi_{2}$ system \cite{VonauPRL05,JPhysD2011}. Here the FT-STS
showed elliptic features around M points which have been
attributed to a fingerprint of large extension of the VHs. As
these results were obtained on the empty states  in  the band
structure, we wanted to explore the filled states  with ARPES
measurements. In the case of the  epitaxied graphene ML on
SiC(0001), the graphene dispersion band is  not contaminated by
the bulk band structure for a large scale of energy and up to the expected VHs \cite{VarchonPRL2007}.\\

\begin{figure}
\includegraphics[width=8cm]{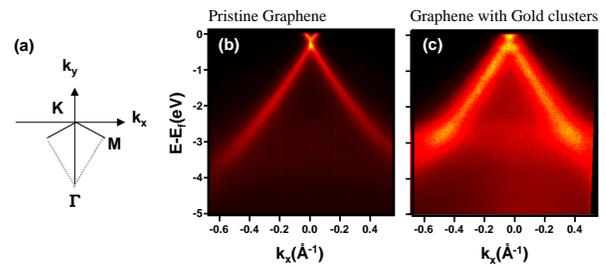}\\
\caption{ARPES intensity of the graphene $\pi$ band around K
points along the direction schematized in a). In b) the ARPES
measurement is done on the pristine ML graphene and in c) on the
graphene ML with intercalated gold
clusters.}\label{Fig2}\end{figure}

\begin{figure}
\includegraphics[width=8cm]{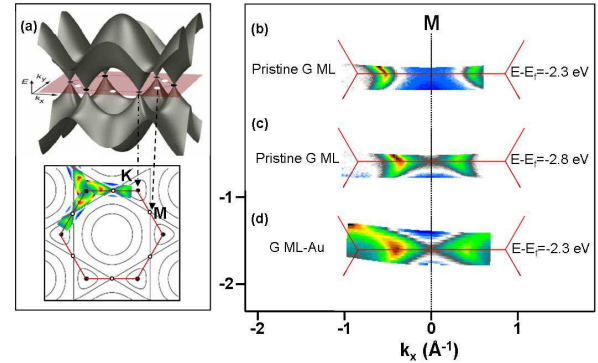}\\
\caption{a) 3D representation of the band structure of graphene
and 2D Constant Energy Contours map (CECs). The CECs are
calculated in a 1NN TB approximation. The key features of the band
structure i.e. the Dirac points and Van Hove singularities (VHs)
are indicated on the 3D representation. b), c) and d, show
equienergetic contours of the photoemission intensity around M
points;   b) and c)for the pristine graphene  at -2.3 eV and -2.8
eV respectively, and d) for the graphene with intercalated gold
clusters at -2.3  eV. The high extension of VHs observed in the
case of graphene ML with intercalated gold cluster in d) is
clearly seen while for the same energy in b) the VHs is not
reached for pristine graphene. The isoenergetical surface obtained
by three-fold symmetrization of the contour showed in d) is
reported in a).}\label{Fig3}\end{figure}

The experimental spectral functions of our epitaxial ML graphene
samples, without and with intercalated gold clusters are shown
respectively in figures \ref{Fig2} b) and c). Both samples exhibit
the characteristic linear dispersion around the K point. The Dirac
point is at 230 meV below the Fermi level for the functionalized
graphene with gold clusters and 260 meV  for the pristine
graphene. The doping due to the clusters is quite negligible as
previously deduced from STS measurements \cite{PremlalAPL09}. For
the pristine ML graphene (Fig. \ref{Fig2}b), we obtain a very good
dispersion characterized by thin bands and linearity over nearly
$2  eV$. The dispersion band indicates that the pristine ML
graphene is of very high quality. In the case of graphene with
intercalated gold clusters (Fig.\ref{Fig2}c) the bands are much
broader. This is probably due to the reduced size of the
homogeneous domains. However the dispersion is also linear until
-2 eV .  The band exhibits a strong "kink" when approaching the
VHs below -2 eV.
 This is associated to a high
increase in the spectral intensity. This is the  characteristic of
a strong renormalization of the band structure which is usually
due to a doping dependence with electron-electron correlation,
electron-phonon coupling or electron-plasmon coupling.  The
surprise here comes from the Fermi velocity of the quasiparticles.
For the pristine graphene we deduced from the slope a group
velocity of $1.02 \pm 0.08\times10^{6} m.s^{-1}$ while $1.2 \pm
0.2\times10^{6} m.s^{-1}$ is measured in the case of the
functionalized graphene with gold. Although the uncertainty is
larger in the latter case, this Fermi velocity is increased by
more than 20\% of the initial value. \\The constant energy
surfaces around the M points give also remarkable results. The
figure \ref{Fig3}a) recalls the band structure and the interesting
topological points of graphene with the calculated Constant Energy
Contours (CEC). In b), c) and d) we present respectively the
constant energy maps around the M point (which is the position of
the expected VHs) at different energies. The VHs are found at -2.8
eV in the case of pristine graphene in figure 3c) and at -2.3   eV
in the case of graphene with gold in fig. \ref{Fig3} d). As shown
in 3b) the VHs is not reached at -2.3 eV without the presence of
gold clusters. The isocontour in d) (with gold) is strongly warped
and the apex of the triangular shaped contour is filled
approaching the VHs. These results confirm our interpretation of
standing waves  pattern and features observed in FT-STS attributed
to a large VHs extension \cite{cranneyEPL10}. The reason of such a
VHs extension remains an open question for the moment. It seems
that there is a general behavior of the graphene band structure in
the case of intercalated compounds (graphite and epitaxial
graphene) particularly around the VHs. Indeed, similar results
have been obtained on epitaxial graphene where strong VHs
extension has been observed after the intercalation of K and Ca in
epitaxial graphene layer \cite{McChesneyPRL2010}. In this latest
experiment the graphene was strongly n-doped and the Fermi level
was positioned at the VHs of the graphene-derived $\pi^{*}$ states
which are usually suspected to be at the origin of the
superconductivity. Up to now, whatever the studied system (GICs or
intercalated graphene), the intercalant was considered to be
homogeneously distributed between graphene planes leading to
specific surstructures (usually a p-2x2) depending on the
stoichiometry\cite{Dresselhauss2002}. However in each of these
cases the VHs extension has been associated to a highly doping
process and electron-electron correlation. The Fermi velocities
were found generally lower (0.5 to $0.7\times10^{6}  m.s^{-1}$).
In our case the electron-electron correlation has to be ruled out.
The sample is no more doped compared to the pristine graphene and
the Fermi velocity is increased by the intercalated clusters. The
Fermi velocities,  we have measured for Pristine graphene and with
intercalated gold clusters  are comparable to those found in the
literature,  i.  e. $0.9 $ to $ 1\times10^{6} m.s^{-1}$ for
epitaxial graphene on SiC for the Si terminated face and
$1.1\times10^{6} m.s^{-1}$  for the carbon terminated face (see
for example \cite{SeyllerPSS2008}), or
 in the case of exfoliated
graphene on $SiO_{2}$\cite{NovoselovNature2005}. Then the Fermi
velocity we have measured with the intercalated gold clusters is
comparable to the one measured on graphene epitaxied on the
C-face. This tends to demonstrate that the clusters decouple the
graphene layer from the substrate as in  the case of C- face
graphene layers. However this leads to a counterintuitive
reasoning. Indeed,  among the all possible origins of the VHs
extension, the pseudo periodic potential created by the clusters
is the most probable hypothesis.
 Following the theoretical calculation of Cheol et al.
\cite{LouieNature2008}, applying a weak pseudopotential should
leads to decrease the group velocity (renormalization). This work
also shows that a slight potential oscillation and/or corrugation
on the graphene layer tends to strongly modify  the VHs. Here the
compressive strain of graphene is  probably partially released by
the decoupling induced by the clusters, leading to an increase of
lattice parameter and consequently an increase of the Fermi
velocity compared to pristine graphene.  Concerning the VHs
extension, one more time this appears to be a general
characteristic of the graphene electronic properties. Indeed,  as
soon as the hopping process of the quasiparticule is modified,
either theoretically by the modification of the Hopping potential
 \cite{BenaPRB11} or experimentally by the modification of the
rotation angle between two graphene layers
\cite{AndreiNatPhys2010}, the VHs are affected. In a very complete
detailed study of the graphitic superconductors, it has been
demonstrated that critical temperature  Tc depends on the charge
transfer between the intercalated atoms and the graphene plane
i.e. the doping level. The increase of Tc was correlated to an
expected electron-phonon coupling which is associated to the kink
in the dispersion band structure \cite{VallaPRL2011}. This was
also associated to a strong anisotropic Fermi velocity and
trigonal warping in the constant energy contour map around the K
point approaching the Fermi level. The reason why these clusters
modify the band structure in this way remains an open question but
in this context, considering the importance of the controversial
possible role of the VHs in the superconductivity property (VHs
extension scenario), particularly in the case of GICs, our results
open a new way for  the understanding of this phenomena. Indeed
here the graphene is not strongly doped, and the intercalant is
not homogeneously distributed. The last but not the least point of
this study is to show one more time the link between the FT-STS
and ARPES measurements, even if it is not discussed in detail
\cite{cranneyEPL10}.

 \acknowledgements This work is supported by the R\'{e}gion Alsace
and the CNRS. The Agence Nationale de la Recherche supports this
work under the ANR Blanc program, reference
ANR-2010-BLAN-1017-ChimiGraphN

\newpage
\end{document}